\documentclass[onecolumn, aps]{revtex4}

\usepackage{latexsym,amsmath,amssymb}

\newcommand{\be}{\begin{eqnarray}}
\newcommand{\ee}{\end{eqnarray}}

\newcommand{\Rar}{\Rightarrow}
\DeclareMathOperator{\sech}{sech}

\DeclareMathOperator{\ctgh}{ctgh}
\newcommand{\exclude}[1]{}

\begin{document}

\title{The cosmological constant as a manifestation of the conformal anomaly?}

\author{Evan C. Thomas, Federico R. Urban, Ariel R. Zhitnitsky}

\affiliation{Department of Physics \& Astronomy, University of British Columbia\\
6224 Agricultural Road, Vancouver, B.C. V6T 1Z1, Canada}

\date{\today}

\begin{abstract}
We propose that the solution to the cosmological vacuum energy puzzle may come from the infrared sector of the effective theory of gravity, where the impact of the trace anomaly is of upmost relevance. We proceed by introducing two auxiliary fields, which are capable of describing a diversity of quantum states via specification of their macroscopic (IR) boundary conditions, in contrast to ultraviolet quantum effects. Our investigation aims at finding a realistic cosmological solution which interprets the observed cosmological constant as a well defined deficit in the zero point energy density of the Universe. The energy density arises from a phase transition, which alters the properties of the quantum ground state.
We explicitly formulate low energy gravity as an effective field theory with a precise definition of the ``point of normalization" as the point at which the ``renormalized cosmological constant" is set to zero in the Minkowski vacuum, in which the Einstein equations are automatically satisfied as the Ricci tensor identically vanishes. With this definition the effective QFT of gravity has a predictive power. In particular, it must predict the evolution of the system in any nontrivial geometry, including the vacuum energy behaviour as a function of infrared, rather than ultraviolet, input parameters.
\end{abstract}

\maketitle

\section{Introduction}\label{intro}
The striking observation that our Universe is accelerating is providing cosmologists and particle theorists with one of the most intricate and puzzling problems in modern physics.  Over the past decade, experimental evidence supporting a non-zero cosmological energy density which appears to be non-clustered and homogeneously and isotropically distributed across the Universe, has grown to almost certainty~\cite{Spergel:2003cb,Riess:1998cb,Perlmutter:1998np} (see also~\cite{Copeland:2006wr,Sarkar:2007cx} for more up-to-date references).  Although there persists room for different explanations~\cite{Mustapha:1998jb,Celerier:1999hp,Tomita:2000jj,Zibin:2008vk}, and the observational data may not be fully solid and should be taken with some precautions~\cite{Sarkar:2007cx}, the so called ``Concordance Model'' (or $\Lambda$CDM for Cold Dark Matter), despite its disquieting implication that we do not know what the great majority of the Universe is made of, is nowadays widely accepted.

Observational results tell us that the Universe is permeated with an unknown form of energy density which makes up for about 75\% of the total energy density, which appears to be exactly the critical ratio for which the three-dimensional spatial curvature is zero.  In numbers this is usually stated as follows
\be\label{lambda-term}
\Omega_\Lambda h^2  \approx 0.39 \, .
\ee

In spite of the mass of models that have been conceived by theorists (see~\cite{Copeland:2006wr} for a comprehensive review), the picture is still unclear, as most models encounter some sort of conceptual or observational obstacles.  It is customary to associate the ``dark'' energy density with vacuum fluctuations, whose energy density would be proportional to the fourth power of the cutoff scale, associated to the highest energy wave modes, at which the underlying theory breaks down.  If this argument were true, we would be faced with a disagreement between theory and observation varying between ~40 to ~120 orders of magnitude.  Clearly, this can not be true, and more complex ideas must be explored.  Notice that most models in the literature adopt the same view, and therefore try to cancel or suppress short distance vacuum fluctations in one way or another~\cite{Copeland:2006wr}.

In the present paper we advocate a fundamentally different viewpoint toward the dark energy / cosmological constant problem.  More precisely, our guiding philosophy is that gravitation, in its present form, can not be a truly fundamental interaction, but rather it is a low energy effective interaction. In such a case, the corresponding gravitons should be treated as quasiparticles which do not feel all microscopic degrees of freedom, but rather are sensitive to the ``relevant excitations" only. We  note that such a viewpoint represents a  standard effective lagrangian approach in all other fields of physics such as condensed matter physics, atomic physics, molecular physics, particle physics. In particular, in condensed matter physics, a typical scale of the problem is  in the eV range, which has nothing to do with the electron mass  which is in the MeV, or the nuclei mass in the GeV. The relevant quasiparticles simply do not know about MeV or GeV scales as in the effective lagrangian approach those scales are effectively tuned away and never enter the system. We should say that this philosophy is neither revolutionary nor new, rather, it has been discussed previously in the literature, see some relatively recent papers~\cite{Bjorken:2001pe,Schutzhold:2002pr,Klinkhamer:2007pe,Klinkhamer:2008nn} and references on previous works therein.

Notice that in general such formulations possess some level of Lorentz non-covariance, see for instance the discussion in~\cite{Bjorken:2001pe}.  To be more precise, in this respect, there are two problems, which could be in principle related: the first of which is that explicitely addressed by Bjorken in~\cite{Bjorken:2001pe}, that is, the appearance of condensates at the Lagrangian level; the second one appears only at the level of the boundary conditions one has to specify in a theory with spacetime boundaries (which so far we have not discussed, but will appear below).  As far as the first kind of non-covariance is concerned, we believe our theory is safe because there are no explicit Lorentz violating condensates at the Lagrangian level.  On the other hand, we certainly have Lorentz violation to a certain degree due to the fact that we need to impose boundary conditions on some infrared auxiliary fields (see below for their precise definitions) and that these in principle are not Lorentz invariant.  This would be the case for instance of a field theory on a compact manifold, but the extent of Lorentz violation in this case should be easily made safe by taking a large enough domain (e.g., larger than the Hubble scale) so to not be in conflict with experiment.

Now, if we accept the framework of the effective quantum field theory for Gravity, than  the basic problem of why the cosmological constant is 120 orders of magnitude smaller than its ``natural" Plank scale $M_{Pl}^4$ is replaced by a fundamentally different problem: what is the relevant scale which enters the effective theory of gravitation? This effective scale obviously has nothing to do with cutoff scale $\sim M_{Pl}$ which is typically associated with the highest energy ultraviolet (UV) scale, at which the underlying theory breaks down. Instead, the relevant effective scale must appear as a result of a subtraction at which some infrared (IR) scale enters the physics. The IR scale may do so in a number of different ways:  through the boundary conditions and/or as a point of normalisation, or by some other means. What is important is that an effective QFT of gravitation has an IR parameter in its definition in contrast with the UV parameter which appears if gravitation is defined as  a truly fundamental theory.

For example, it has been argued in~\cite{Klinkhamer:2007pe} that in the effective QFT approach the Gibbs free energy (rather than the original total energy) plays the role of macroscopic vacuum energy interacting with gravitation. It has been also argued that it vanishes in the absence of external pressure. However, the logic of the effective QFT neither addresses nor requires an explanation for the relative smallness of $\Omega_\Lambda$ in comparison with ``natural'' Plank scale $M_{PL}^4$:  one can always  introduce the bare cosmological constant to absorb the corresponding UV divergences into the energy density in such a framework. It can be done however only once and forever and only at a chosen specific ``point of normalisation''. After this procedure is performed the dark energy (and all other relevant observables) must be derived from the effective QFT describing Gravity at low energy scales.

According to this logic, it is quite natural to define the ``renormalised cosmological constant'' to be zero in Minkowski vacuum with metric $\eta_{\mu\nu}= \text{diag} (1,-1,-1,-1)$ wherein the Einstein equations are automatically satisfied as the Ricci tensor identically vanishes in flat space. Thus, the energy momentum tensor  $\langle T_{\mu\nu} \rangle \sim \eta_{\mu\nu}$ in combination with this ``bare cosmological constant'' must also vanish at the specific ``point of normalisation'' to satisfy the Einstein equations. Once this procedure is performed, the effective QFT of gravitation must predict the behaviour of the system in any nontrivial geometry of the space time. From this definition it is quite obvious that the ``renormalised energy density'' must be proportional to the deviation from the flat space time geometry, for example to the Hubble constant $H$ in case of  the de Sitter metric, $a(t) =\exp (Ht)$. From this definition it is also obvious that all dimensional parameters, such as masses of particles and fields which contribute to the trace of the energy momentum tensor $\langle T_{\mu}^{\mu} \rangle$ in the Minkowski vacuum, must cancel with the ``bare cosmological constant" after an appropriate subtraction procedure resulting in zero vacuum energy in Minkowski vacuum. This statement remains valid  for classically non-vanishing contributions to the trace of the energy momentum tensor $\langle T_{\mu}^{\mu} \rangle$ (due to massive particles) as well as quantum anomalous contribution such as nonzero gluon condensate in flat space. Stated differently, a nonzero contribution to the energy density emerges only as a result of deviation from flatness, and therefore must be proportional to some (positive) power of  $H$ in case of  the de Sitter metric. Of course, our framework does not, can not, and should not answer the ``fundamental" question of the relative size of the cosmological constant. This kind of question simply does not arise in the effective QFT formulation. Still, the gravity being treated as the effective quantum field theory does have predictive power: it addresses a number of other questions formulated in terms of the IR parameters which enter the system  at the specific ``point of normalisation''. For example, it gives a low-energy expression for the cosmological constant.

The conformal anomaly, being an IR contribution to the stress-energy, plays a crucial role in understanding of behaviour of  the energy momentum tensor $\langle T_{\mu}^{\mu} \rangle$ at low energy in a curved space. In the case of free particles the corresponding computations have been carried out long ago~\cite{Birrell:1982ix} with a typical result, see below for details,
\be
\label{H^4}
\langle T_{\mu}^{\mu} \rangle \sim H^4.
\ee
This is an astonishingly small number which can be ignored for all imaginable applications. However, this is not a final sentence for the effective QFT approach to gravitation. In fact, there are a number of arguments suggesting that the interactions can drastically change
this naive estimate (\ref{H^4}) especially if a nonlocal effective interaction  (corresponding to an induced  long distance interaction) emerges. In particular, in refs.~\cite{Bjorken:2001pe,Schutzhold:2002pr,Klinkhamer:2007pe,Klinkhamer:2008nn} it is argued that 
\be
\label{H}
\langle T_{\mu}^{\mu} \rangle \sim H \Lambda_{QCD}^3 \sim \left(10^{-3}  \text{eV}\right)^4
\ee
instead of $\langle T_{\mu}^{\mu} \rangle \sim H^4$. The estimate (\ref{H}) is amazingly close to the observed vacuum energy density today.  It is quite instructive that $ \Lambda_{QCD} $ appears in the problem. Indeed, QCD is the only well-understood fundamental strongly-interacting QFT realised in nature. The electroweak  theory is actually weakly coupled, therefore, it is unlikely that the corresponding almost non-interacting heavy degrees of freedom contribute to the vacuum energy with the definition for the vacuum energy formulated above. In fact, a quite  general argument  has been presented~\cite{Schutzhold:2002pr} suggesting that the first power in $H$ similar to eq.(\ref{H}) may emerge only for sufficiently strongly-interacting QFT, while 
a most likely outcome of the computations for non-interacting or weakly interacting theories  will be  given by (\ref{H^4}), and therefore will be  irrelevant for any applications~\footnote{Notice that covariance arguments demand that only even powers of $H$ appear in the expression for the vacuum energy~\cite{Shapiro:2008sf}; this simple dimensional argument is not applicable in some circumstances such as a non trivial global topology as explicit  computations in 2D toy-model shows~\cite{Urban:2009wb}. The simplest way to explain this feature of nontrivial topology is to introduce a temperature which can be implemented by considering the topology of a cylinder.  As is well known, one should generically    expect any power of temperature $T$ in the expansion for the vacuum energy at small $T=1/L$.  If the size $L$ of the system scales as $L\sim H L_0$~\cite{starobinsky} it would automatically lead to the desired scaling.}.

It is also quite encouraging that the well-known scale $\Lambda_{QCD}\sim 100 $ MeV apparently is precisely what is required
 to understand the observational  vacuum energy density  $ \left(10^{-3}  \text{eV}\right)^4 $ if one assumes that $\langle T_{\mu}^{\mu} \rangle \sim H \Lambda_{QCD}^3$ \footnote{ It is also quite amazing that QCD may play a key role in explaining a number of dark matter puzzles along with the dark energy mystery which is the subject of the present study, see recent paper~\cite{Forbes:2008uf} and references therein. The emergence of the QCD scale in the vacuum energy may finally unlock the mystery of the ``Cosmic Cincidence'' problem, that is the observational similarity among visible and invisible scales: $\Omega_{\Lambda} \simeq 4 \Omega_{DM}$ and $\Omega_{DM } \simeq 5 \Omega_{B}$ where only $\Omega_{B}$ (which represents the baryon contribution to $\Omega_{tot}=1$) has an obvious relation to QCD, as the nucleon mass $m_N$ is proportional to $\Lambda_{QCD}$ while ``God's particle" --the Higgs boson, contributes only a few percent to $m_N$ through the quark's mass proportionality to the vacuum expectation value of the Higgs field. }. It is also interesting to note that a similar estimate was given in 1967  by Zeldovich~\cite{Zeldovich:1967gd} who argued that  $\langle T_{\mu}^{\mu} \rangle \sim {m_p^6}/{M_{PL}^2}$ which is precisely the estimate (\ref{H})   if one replaces $ \Lambda_{QCD} \rightarrow m_p $ ($ \Lambda_{QCD} $ was not known at that time) and $  H\rightarrow \Lambda_{QCD}^3/ M_{PL}^{2} $~\footnote{See~\cite{Sahni:1998at} for another application of gravity IR effects in cosmology.  References~\cite{Elizalde:1994av,Shapiro:2004ch,Sola:2007sv} investigate the effects of a running in the vacuum energy term; an approach working directly with the equations of motion for gravity is put forward in~\cite{Bauer:2009ke}.}
 
The main goal of the present  paper is a systematic  analysis  of the anomalous contribution to the  vacuum energy density  $ \langle T_{\mu}^{\mu} \rangle $ to see under what conditions the conformal anomaly can provide an appropriate  scale observed in nature  $ \langle T_{\mu}^{\mu} \rangle \sim H \Lambda_{QCD}^3$. We also want to study the cases in which such a scale does not emerge (negative lessons are also lessons).  Fortunately, the technical tools to attack this problem have recently been developed~\cite{Mottola:2006ew,Antoniadis:2006wq,Anderson:2007eu,Giannotti:2008cv}. Therefore, we shall study the behaviour of $ \langle T_{\mu}^{\mu} \rangle$ as a result of a conformally anomalous contribution with some specific non-local induced interactions as argued in~\cite{Mottola:2006ew,Antoniadis:2006wq,Anderson:2007eu,Giannotti:2008cv} (See also~\cite{Odintsov:1991tg,Shapiro:1994ww,Elizalde:1994nz,Balbinot:1999ri,Balbinot:1999vg}).

\section{The conformal anomaly} 
As mentioned in the introduction, the problem has recently been tackled from a different point of view, namely focussing on the low energy band of the spectrum of General Relativity (GR), viewed as a low energy effective quantum field theory (QFT), in some analogy with the view taken in dealing with the Casimir effect~\cite{Mottola:2006ew} (see also~\cite{Fulling:2007xa,Milton:2007ar,Shajesh:2007sc} for the gravitating properties of the Casimir energy).  In this approach  GR is augmented with two (in four dimensions) additional auxiliary higher derivative scalar fields, which are most relevant at large distances (regarded as an IR completion of GR, instead of a high frequency UV one).  The new fields contribute through their Lagrangian to the low energy effective semiclassical description of GR, and their energy momentum tensor reproduces exactly the conformal anomalous terms, which arise necessarily once GR is treated as a semiclassical QFT.

It is natural to ask that this model, as a complete low energy effective theory of quantum gravity, to provide an explanation for the Dark Energy (DE) puzzle, as a phenomenon related solely to IR physics and the conformal anomaly; this is precisely the point we address here.  First, a brief review of the ideas pushed forward in the paper~\cite{Mottola:2006ew} is given;  secondly, the details of the physics which may be responsible for the non-compensation of vacuum energy observed today are outlined, with special emphasis on the role of the auxiliary fields;  the summary of our results will follow, together with an overview on the forthcoming steps in the direction of a more realistic and complete analysis.  Another interesting proposal, namely a secular screening of the cosmological constant, has been investigated in~\cite{Prokopec:2002jn,Prokopec:2002uw} (See~\cite{Starobinsky:1980te} for anomaly-driven inflation).

\subsection{What is the conformal anomaly?}\label{confanom}
In QFT an anomaly is a classical symmetry which is broken by quantum effects, that is, as the classical theory is quantised.  Minkowski QFT's are populated with such anomalies, related to the non trivial vacuum configurations of a particular theory.  In curved space the situation is somewhat more complicated, but it is now well estabilished that once classical GR is considered as a background for a conformally coupled QFT, the classical expectation that the trace of the matter fields energy momentum tensor vanishes is incorrect. Indeed, when the energy momentum tensor is treated as a quantum expectation value, more terms in the effective lagrangian arise during the process of renormalisation, which in turn lead to additional terms in the Einsten equations (EE).  In particular, the conformal (or trace) anomaly in four dimensions takes the form
\be\label{anomaly}
\langle T_\mu^{\ \mu} \rangle = b F + b' \left(E - \frac{2}{3} \Box R \right) + b'' \Box R \, ,
\ee
where
\be
E &\equiv& ^*\! R_{\mu\nu\rho\sigma} \!\, ^*\! R^{\mu\nu\rho\sigma} = R_{\mu\nu\rho\sigma}R^{\mu\nu\rho\sigma} - 4 R_{\mu\nu}R^{\mu\nu} + R^2 \, , \label{escalar} \\
F &\equiv& C_{\mu\nu\rho\sigma}C^{\mu\nu\rho\sigma} = R_{\mu\nu\rho\sigma}R^{\mu\nu\rho\sigma} - 2 R_{\mu\nu}R^{\mu\nu}  + \frac{R^2}{3} \, . \label{fscalar}
\ee

We follow the conventions ( - + + +) in the notation of~\cite{Misner:1974qy}, and denote $^*\! R_{\mu\nu\rho\sigma}= \varepsilon_{\mu\nu\lambda\kappa}R^{\lambda\kappa}_{\ \ \rho\sigma}/2$ the dual of the Riemann curvatire tensor $R_{\mu\nu\rho\sigma}$;  $C_{\mu\nu\rho\sigma}$ is the Weyl conformal tensor.  The coefficients $b$, $b'$ and $b''$ are dimensionless parameters which depend on the number of free massless fields contributing to the anomaly.  It is important (as we shall see later) that only free and effectively massless are counted in when computing these coefficients.  The form of (\ref{anomaly}) and the coefficients $b$ and $b'$ do not depend on the state in which the expectation value of the stress tensor is computed.  Indeed, they depend only on the number of massless fields~\cite{Birrell:1982ix} through the formulas
\be
b &=& - \frac{1}{120 (4 \pi)^2} \, (N_S + 6 N_F + 12 N_V) \, , \nonumber\\
&& \label{bcoeff} \\
b'&=& \frac{1}{360 (4 \pi)^2} \, (N_S + 11 N_F + 62 N_V) \, , \nonumber
\ee
with $(N_S, N_F, N_V)$ the number of scalar, fermion, and vector degrees of freedom contributing to the anomaly (notice the sign difference with respect to~\cite{Mottola:2006ew}).  The origin of the third coefficient $b''$ is still unclear, as it is possible to redefine this coefficient by adding a local counterterm proportional to $R^2$ in the gravitational action.  It appears then that this coefficient is regularisation dependent, and could be possibly removed by a suitable gauge choice. As such we do not consider it a part of the true anomaly.
On the other hand, the $b$ and $b'$ terms can not be derived by a local action containing only the metric and its derivatives, unless this action is singular~\cite{Birrell:1982ix,Koksma:2008jn}.  However, an appropriate non-local action, which will therefore contain information about large scales, can be constructed and is finite, as we will see shortly. This non-local action can be expressed as a local action with the introduction of the previously mentioned auxiliary fields.

\subsection{The dynamics of the conformal factor}\label{conffac}
To explicitely see how a non-local action incorporating the conformal anomaly is put together, we must look at the dynamics of the conformal degrees of freedom $\sigma(x)$ for conformally related metrics
\be\label{gconf}
g_{\mu\nu} = e^{2\sigma} \, \bar g_{\mu\nu} \, .
\ee

Proceedings along the lines of~\cite{Mottola:2006ew}, one constructs the operators
\be
\sqrt{-g} \, F &=& \sqrt{-\bar g} \, \bar F \, \label{operatorf} \\
\sqrt{-g} \, \left(E - \frac{2}{3} \Box R \right) &=& \sqrt{-\bar g} \, \left(\bar E - \frac{2}{3} \bar{\bar\Box} \bar R \right) + 4 \, \sqrt{-\bar g} \, \bar\Delta_4 \, \sigma \, , \label{operatore}
\ee
where the fourth order differential operator (which is conformally covariant in four dimensions) is given by
\be\label{delta4}
\Delta_4 \equiv \Box^2 + 2 R^{\mu\nu} D_\mu D_\nu - \frac{2}{3} R \Box + \frac{1}{3} (D^\mu R) D_\mu \, ,
\ee
with $\Box=D_\mu D^\mu$ and $D_\mu$ the usual covariant derivative.  It is now straightforward to write down the non-local action which would, upon differentiation with respect to the conformal factor, return the conformal anomaly (\ref{anomaly}).  This effective action is given by
\be\label{wz}
\Gamma[ \bar g ; \sigma] = b \int \, d^4x \, \sqrt{- \bar g} \, \bar F \, \sigma + b' \int \, d^4x \, \sqrt{- \bar g} \, \left\{ \left(\bar E - \frac{2}{3} \bar{\bar\Box} \bar R \right) \sigma + 2 \, \sigma \bar \Delta_4 \sigma \right\} \, .
\ee
It is noteworthy how this effective action is a function of the conformal factor and the metric $\bar g_{\mu\nu}$, because it is impossible to find a corresponding action function only of the original metric $g_{\mu\nu}$.  This fact indicates how, in order to describe the conformal anomaly, we must be considering additional degrees of freedom, in this case represented by the conformal factor.

The conformal anomaly is derived, in this language, as a conformal variation of the (Wess-Zumino) action (\ref{wz}).  Furthermore, in this formalism, it is also possible to formally remove the conformal factor by solving (\ref{operatorf}, \ref{operatore}), and recognising that
\be\label{deltawz}
\Gamma [\bar g ; \sigma] = S_A [g = e^{2 \sigma} \bar g] - S_A [\bar g] \, ,
\ee
where the non-local anomalous action is expressed as
\be\label{anomaction}
S_A [g] =  \frac{1}{8} \int d^4x \sqrt{-g} \int d^4x' \sqrt{-g'} \left[E - \frac{2}{3} \Box R \right]_x \Delta_4^{-1} (x,x') \left[ 2bF + b' \left(E - \frac{2}{3} \Box R \right) \right]_{x'} \, .
\ee

The next step will be to render this action local thanks to the introduction of two (in four dimensions) auxiliary scalars, encoding the information now stored in (\ref{wz}).

\subsection{The infrared auxiliary fields}\label{auxfields}
The anomalous action (\ref{anomaction}) is considered as a fundamental ingredient of the low energy effective theory of (quantum) gravity, and as such it will contribute to the total stress tensor of the system.  In order to see how this comes about  explicitly, one is prompted to write the non-local action in terms of a local action which will contain extra degrees of freedom.  This is most easily done by introducing two auxiliary scalar fields obeying the Euler-Lagrange equations
\be
\Delta_4 \, \varphi &=& \frac{1}{2} \left(E - \frac{2}{3} \Box R \right) \, , \label{auxphi} \\
\Delta_4 \, \psi &=& \frac{1}{2} F \, , \label{auxpsi}
\ee
in terms of which the anomalous action (\ref{anomaction}) becomes
\be\label{auxaction}
S_A [g ; \varphi , \psi] &=&  \frac{b'}{2} \, \int \, d^4x \, \sqrt{-g} \, \left\{ - \varphi \Delta_4 \varphi + \left(E - \frac{2}{3} \Box R \right) \varphi \right\} \nonumber\\
&+&  \frac{b}{2} \, \int \, d^4x \, \sqrt{-g} \, \left\{ - \psi\Delta_4\varphi - \varphi \Delta_4 \psi + F \varphi + \left(E - \frac{2}{3} \Box R \right) \psi \right\} \, ,
\ee
which is the starting point for the practical calculations we will be carrying out.  Notice how the action became entirely local, and the dependence on the confomal factor (as well as on the conformally related metric $\bar g_{\mu\nu}$) dropped out.  This is the true local action corresponding to the anomaly in four dimensions.

The total action for the low energy theory is therefore
\be\label{totaction}
S_{\textrm{tot}} = S_{\textrm{EH}} + S_{\textrm{Weyl}}^{(2)} + \sum_{n\geq3}^\infty S_{\textrm{local}}^{(n)} + S_A \, ,
\ee
where the first term is the usual Einstein-Hilbert action;  the second term is the conformally invariant Weyl action;  the sum appearing in the third place includes all possible higher order curvature invariants, suppressed by $(2n-4)$ powers of $M_{\textrm{Pl}}$ at low energy;  the last piece is the anomalous contribution.

As usual, there is a corresponding energy momentum tensor, which can be obtained straightforwardly by functional differentiating with respect to the metric, that is
\be\label{emtA}
T_{\mu\nu}^A \equiv - \frac{2}{\sqrt{-g}} \frac{\delta S_A}{\delta g^{\mu\nu}} = b' E_{\mu\nu} + b F_{\mu\nu} \, ,
\ee
where the explicit forms for the two anomalous tensors $E_{\mu\nu}$ and $F_{\mu\nu}$ are given in~\cite{Mottola:2006ew}.  As it should, the trace of this energy momentum tensor is proportional to the equations of motion for the auxiliary fields $\varphi , \, \psi$ which, once employed, return exactly the (on-shell) conformal anomaly
\be\label{traceA}
T^{A \mu}_\mu = 2 b \Delta_4 \psi + 2 b' \Delta_4 \varphi = b F + b' \left(E - \frac{2}{3} \Box R \right) \, .
\ee
This expression corresponds to equation (\ref{anomaly}), but it is obtained using the auxiliary field formalism rather than directly renormalising the gravitational action, or the resulting divergent stress tensor as in \cite{Birrell:1982ix}.
  
Although in the trace of the anomalous stress tensor, the auxiliary fields appear as their equations of motion, and therefore do not appear in the final expression for this trace, it is not so when one singles out the stress tensor components separately.  Indeed, the auxiliary fields will appear in some combinations which will not be removed by the application of their equations of motion.  The energy momentum tensor in general is dependent on the values of the fields $\varphi , \, \psi$, and hence on the particular solutions of their Euler-Lagrange equations (\ref{auxphi}, \ref{auxpsi}).  In particular, the choice of the boundary conditions for the auxiliary fields, in solving their differential equations of motion, will pick out one state from all possibilities, and will explicitly enter in the general expression for the stress tensor, together with the two coefficients $b , \, b'$.

We eventually come up with a correspondence ``dictionary" that, under certain circumstances, enables us to represent the renormalised energy momenum tensor of a given quantum state by choosing a set of boundary conditions for the IR auxiliary fields.

\subsection{Horizon divergencies}\label{horizons}
Let us focus on the general properties of the anomalous stress tensor (\ref{emtA}).  As already noted in~\cite{Mottola:2006ew}, at low energy (that is, in the infrared) the only terms that are relevant in the action (\ref{totaction}) are the usual Einstein-Hilbert one, the conformally invariant Weyl action, and the anomalous part.  In general the anomalous contribution to the stress tensor and the Weyl one would be of the same order of magnitude, as they both contain four derivatives of the metric.  Moreover, the anomalous action is clearly not uniquely defined by (\ref{deltawz}), because any conformally invariant term can be added to it, resulting in the same equations of motion.  However, if the spacetime under examination is in addition conformally flat, then the Weyl invariant vanishes, leaving the anomalous contribution as the only additional (and unique) contribution to the energy tensor.  In this case the anomalous stress tensor can be shown to provide the information about the tracefree part of the complete $T_{\mu\nu}$.

The key realisation, at this point, is that any spacetime with a static Killing horizon is conformally related to flat spacetime near this horizon, and therefore the energy momentum tensor would be accurately approximated by the sole contribution coming from the anomalous action.  Hence, we will be working with spacetimes with horizons, and in particular with de Sitter spacetime in static coordinates (or closely related coordinates systems) which is also conformally flat.

In general, when dealing with spacetimes with horizons, one expects the solutions of the Euler-Lagrange equations for any field in this background to diverge at the coordinate singularity.  This fact reflects on the behaviour of the stress tensor near the horizon, which should inherit the singular behaviour of the fields from which it originated.  This is indeed the case for a number of examples in QFT in curved spacetimes, for instance, the Boulware state in de Sitter space, or in Schwarzschild space.  Notice that this fact does not necessarily imply that the quantum state is to be rejected as non-physical, see~\cite{Mottola:2006ew}.

This singular behaviour signals the breakdown of the semiclassical approximation used throughout the computation leading to the renormalised stress tensor, because of the enormous backreaction on the background geometry.  The divergence as it is must therefore be removed, if we are to trust our low energy effective lagrangian in that region of space bordering the horizon.  The original proposal~\cite{Mottola:2006ew} involves the construction of a static system in which the metric flips to another allowed solution of the EE before reaching the horizon, and flips once more once the horizon (masked by the use of a non-singular metric in that region) has been crossed.

Our proposal goes in a different direction, and makes use of the fact that during the evolution of the Universe matter fields change their properties, such as their (effective) mass, thereby changing their contribution to the anomalous action (again, we will see that only free and massless fields participate in the conformal anomaly).  Consequently, the stress tensor associated to the anomaly went through some kind of ``phase transitions'' as the Universe grew older.  With this mindset at hand, it is natural to ask the question of whether the divergent pieces of $T^A_{\mu\nu}$ remained the same, and if they did, what happens instead to the finite contributions.  In this picture the observed (vacuum) energy would be exactly this finite differential obtained by subtracting the stress tensor after the transition from the one before it, thereby removing the divergencies, but leaving a finite, anomalous, non-clustered, homogeneously and isotropically distributed, (dark) energy density. This finite term is originates precisely at the moment when the spectrum of particles contributing to the anomaly drastically changes. The scale at which this happens should then be the IR scale of the problem. Our goal is thus to see how an internal QFT scale (such as $\Lambda_{QCD}$) manifests itself in the final expression for  (dark) energy density.

\section{Dealing with the divergencies}\label{dealing}
In this section we formulate more precisely the idea that has been previously hinted to, that is, that the anomalous energy momentum tensor underwent some transitions during the evolution of the Universe, due to the change in the number of fields participating in the anomaly, which is ultimately due to their properties being modified by the interaction with other fields.  We will be discussing the background physics for these ideas, indicating whether their effects are believed to be significant. The two main examples we will be considering are: first, phase transitions such as QCD, Electro-Weak (EW), and possibly others --e.g.\ Peccei-Quinn (PQ) symmetry breaking-- where some of the fields masses are different at the two boundaries of the phase transition;  and second, the decoupling of species, for in that case, before the actual decoupling, there is an effective mass due to the interaction with the hot plasma, even for formally massless fields like photons.

\subsection{Phase transitions}\label{pht}
As previously stated, the first place where the general arguments given in section \ref{horizons} are likely to play a role is during a phase transition.  This is because, as stated, only effectively massless fields take part in the conformal anomaly, shown by the expressions (\ref{anomaly}-\ref{bcoeff}).
Now consider a phase transition during the course of which some fields change mass, having different values on either side of the boundary of the transition.  This is known to be the case for a number of examples in the early Universe, for instance the EW phase transition giving masses to, in its simplest version, all Standard-Model (SM) particles;  or the QCD phase transition, which exhibits a very peculiar time dependence for the axion mass (as well as several other hadronic states);  or finally, speaking of axions, the breaking of the PQ symmetry which gives birth to an effectively massless axion.

\vspace{0.5cm}
\noindent
{\it \underline{The Electro-Weak phase transition. $T\sim 100$ GeV.}}

At first sight, the EW phase transition may seem an ideal playground for the ideas presented here.  However, the conformal anomaly can not possibly have any observable effect, as long as only SM fields are considered.  The reason for this is that all the fields which gain a mass thanks to the interaction with the Higgs boson reaching the bottom of the mexican hat potential possess electroweak and/or colour charges, which in turn implies that they can not be treated as free field in the early Universe, and will therefore not contribute to the degrees of freedom count in (\ref{bcoeff}), both before and after the transition.  This is because in the derivation of the conformal anomaly one employs free propagators, and it is sensible to expect that interacting fields would not participate in the anomaly, as massive fields do not.

To make this statement more precise, one can rephrase it as follows.  If a particle has a massless pole (and is therefore a candidate ``anomalous'' particle), and this pole remains massless once interactions are turned on, then this particle will effectively be part of the anomaly; if interactions shift the pole however, then the corresponding field drops from the count of degrees of freedom participating in the anomaly.

To understand precisely how this comes about one should recall how the anomalous terms (\ref{anomaly}-\ref{bcoeff}) were derived. The anomaly can be understood as a result of regularisation of the quantum theory when the regulator fields $\phi_R $ with mass $M_R$ are introduced. The statement that the conformal symmetry is explicitly broken on the quantum level is equivalent to the statement that there is a finite contribution (\ref{anomaly}-\ref{bcoeff}) after taking the limit $M_R\rightarrow \infty$. When a field $\phi$ comes with a non-vanishing mass $m_{\phi}$, the trace of the stress tensor receives a canonical contribution (see~\cite{Decanini:2005eg} for a derivation of the anomalous term for a massive scalar field in curved space) $\langle T_\mu^{\ \mu} \rangle\sim m_{\phi}^2\langle \phi^2 \rangle$ along with the corresponding anomalous contribution (\ref{anomaly}-\ref{bcoeff}) which results from the regulator field $\phi_R $.  Now, if we take $m_{\phi}$ to be very large, the corresponding canonical contribution exactly cancels the anomalous one. When $m_{\phi}$ is large but finite, the resulting suppression is of order $\sim R/m_{\phi}^2$ which can be safely neglected for our purposes. From this discussion it must be that the coefficients (\ref{bcoeff})
count massless degrees of freedom, and heavy degrees of freedom are decoupled from low energy physics.  From the same discussions it must be also that the relevant scale which distinguishes massless from effectively massive degrees of freedom is determined by the ratio $\sim R/m_{\phi}^2$.

Precisely the same pattern is also realised for the chiral anomaly when a quark $Q$ has a non-vanishing quark mass $m_Q$, see the appendix of ref.~\cite{Halperin:1997as} for the details. Indeed, when a Q-quark is strictly massless the corresponding axial current has the standard expression for anomaly originated from the regulator field,  $\langle\partial_{\mu}J^{5\mu}\rangle=\langle\frac{\alpha_s}{4\pi}F\tilde{F}\rangle$ where $F$ is gluon field. When the Q-quark becomes massive the divergence of the axial current receives a canonical contribution along with anomalous contribution stemming from the regulator field, $\langle\partial_{\mu}J^{5\mu}\rangle=\langle2m_Q\bar{Q}i\gamma_5 Q\rangle+\langle\frac{\alpha_s}{4\pi}F\tilde{F}\rangle$. Assuming that $m_Q$ is sufficiently large,
$m_Q^2\gg F\sim \Lambda_{QCD}^2$ one can expand the canonical term to observe that the leading term of this expansion exactly cancels with the anomalous term, while all corrections are suppressed by an addition factor $\sim gF/m_Q^2$ such that, $\langle\partial_{\mu}J^{5\mu}\rangle \sim m_Q^{-2}\langle g^3F^3 \rangle +O(1/m_{Q}^4)$.

\vspace{0.5cm}
\noindent
{\it  \underline{The Peccei-Quinn symmetry breaking transition. $T\sim 10^{12}$ GeV.}}

The axion field has been introduced by Peccei and Quinn in order to solve the ``strong" ${\cal CP}$ problem
which remains one of the most outstanding puzzles of the Standard Model, see the original papers~\cite{axion},~\cite{KSVZ},~\cite{DFSZ}, and the recent review~\cite{vanBibber:2006rb}. Thirty years after the axion was invented, it is still considered as a viable solution of the strong ${\cal CP}$ problem, and one of the plausible dark matter candidates. At present, there are several experimental groups searching for the axions, see~\cite{vanBibber:2006rb} and references therein.

The main characteristic of the mechanism proposed by Peccei and Quinn which is of interest to us is the introduction of a new scalar degree of freedom, the axion, which comes to life once the PQ symmetry is spontaneously broken.

At very high temperatures the axion feels no potential, and it is effectively massless.
Therefore, it obviously contributes to the anomaly.   However, due to its interactions with other particles, it can be considered as ``anomalous'' only for a limited time range, as it will become massive as temperature drops,  see below.

\vspace{0.5cm}
\noindent
{\it  \underline{ The QCD confinement and chiral phase transitions. $T\sim 170$ MeV.}}

We should note that it is believed that confinement-deconfinement transition is really a cross-over for the realistic quark masses rather than a true phase transition. Also, it is still unknown whether confinement-deconfinement and the chiral phase transitions happen at exactly the same temperature or at slightly different temperatures. However, for our discussions below this kind of subtlety does not play any role.

The crucial point is that fields are in fact changing properties around this temperature. In particular, the axion field could come into play once more at the moment of the quark/gluon-hadron phase transition. Indeed, the cosmological history of this field is extremely non-trivial around the QCD  epoch, the main feature of which being the time dependence of the axion mass.

At the QCD phase transition, the axion field acquires a small mass (which increases while the temperature decreases) due to the appearance of a periodic potential induced by instantons. Once the temperature drops below the critical value and the chiral condensate forms, the axion mass $m_a$ assumes its final form, $m_a^2f_{PQ}^2\sim m_q\langle \bar{q}q\rangle$.   Moreover, the axion is a scalar field with no electric or colour charge, and tiny cross sections when interacting with SM particles; this makes it an ideal candidate for the exploitation of our arguments.

Hence, the ``effectively massless" axion becomes an ``effectively massive" field at time $t_0$ determined by the condition $R/m_a^2(t_0)\sim 1$.  Thus, we arrive at a scenario in which in the initial phase (unconfined phase) we have an effectively massless and almost free field (the axion) living in some background quantum vacuum state, and in the final phase (confined phase) we have instead a massive axion field, and a different vacuum configuration.  Both these properties should be tracked in the anomalous stress tensor given in (\ref{emtA}), via the change in the coefficients (\ref{bcoeff}) and in the boundary conditions for the auxiliary fields.

Phase transitions may not be the only important consideration in the anomalous stress tensor evolution, since at the same time the vacuum structure undergoes a profound re-organisation, which in turn results again in a corresponding necessary adjustment in the anomalous $T^A_{\mu\nu}$, and in particular in the choice of the boundary conditions for the auxiliary fields (\ref{auxphi}, \ref{auxpsi}).

\subsection{Decoupling of species}\label{dec}
The second generic  situation in which the anomalous stress tensor could experience a sort of transition from one state to another is the decoupling of some species from the equilibrium thermal bath in which, at early times, most fields are immersed.

Let us consider, as an example, the cosmological history of photons.  Charged particles are not of any interest here: as before, charged particles, even when decouped from the thermal plasma, have a mean free path which is shorter than the Hubble length during a given epoch, and therefore can not be treated as free fields at these distances. This is because when calculating the anomaly we calculate some one-loop diagrams using free propagators of various fields going around the loop. If however the mean free path for a particle is shorter than the length scale at which the integrals in the one-loop diagrams are saturated, then we cannot legitimately carry out the calculations using the free propagator for that particle. The result of this is, if a field can not be treated as free, it then can not be part of the anomaly, as long as the anomaly has effects on these scales (which is in fact what we are looking for).

A photon in the early Universe, that is, before decoupling around the last scattering epoch, is in close thermal contact with the other SM particles that have not yet fallen out of kinetic equilibrium.  This translates to the fact that photons will have a small mean free path, which would depend on their momentum, and therefore makes them unsuitable to be considered as free fields contributing to the anomaly.  Things change once photons fall out of kinetic equilibrium, becoming in every sense free and massless.  These photons would necessarily be included in the degrees of freedom count for (\ref{bcoeff}), while the coupled photons would not be counted.  Hence, we again find ourselves in the situation in which a field (at least, for a given frequency range) takes part to the anomaly only on one side of a transition (in this case, kinetic decoupling).

This argument may apply, again, to the axion field, which, depending on its properties (mainly on the value of the PQ scale $f_{PQ}$) could be brought to thermal equilibrium in the early Universe.  Yet another possibility are neutrinos, for they have minuscule masses and cross sections, and undergo a similar decoupling transition at temperatures in the MeV range.

\section{Method, and results}\label{results}
Having laid the physical basis for our arguments, let us now be more specific and analyse in detail the behaviour of the stress tensor in different setups.  The discussion will be kept as general as possible, not restricting ourselves to one specific scenario among those mentioned above.  One more motivation for proceeding this way is that, although expected on general grounds, there is no method available at the moment which can univocally relate the semiclassical vacuum expectation values for such configurations, with the auxiliary fields stress tensor with different boundary conditions.

First of all, let us explicitely write down the set of coordinate systems we will be employing.  As already clarified, our Minkowski metric is $\eta_{\mu\nu} = \textrm{diag}(1,-1,-1,-1)$, and, given that we will be always in de Sitter space, the Ricci tensor and scalar will be always given by
\be\label{ricci}
R_{\mu\nu} &=& \frac{1}{4} g_{\mu\nu} R \, , \nonumber\\
R &=& - 12 H^2 \, ,
\ee
where $H$ is the constant Hubble parameter.  Furthermore, we will need the fourth order curvature invariants $E$ and $F$, which for de Sitter space are given by
\be\label{eandf}
E = 24 H^4 \qquad \textrm{and} \qquad F = 0 \, .
\ee

The coordinate systems are some of those typically employed in de Sitter (dS) spacetime, namely the spatially flat Friedman-Robertson-Walker (FRW) system, the conformally flat (CF) one, the Tortoise system (from the name given to the radial ``tortoise'' coordinate), the static coordinates system, and finally a system in which de Sitter spacetime is conformal to Rindler four dimensional spacetime, which we will call the Rindler system.  Notice that in all coordinate systems we will tamper with the ``radial'' and the ``time'' coordinates, preserving the ${\cal O}(3)$ symmetry of three dimensional rotations.  We refer to the appendix for all the reference formulas.

\subsection{The anomalous $T^A_{\mu\nu}$ in de Sitters static coordinates}\label{emtAexp}
Let us choose, as a reference metric, the static de Sitter one, that is
\be\label{Static-dS}
ds^2 = (1 - H^2 r^2) dt^2 - \frac{1}{1 - H^2 r^2} dr^2 - r^2 d\Omega^2 \, .
\ee

Having solved, in some special cases, the equations of motion for the auxiliary fields (see the appendix), it is now possible to have explicit expressions for the anomalous $T^A_{\mu\nu}$ in terms of the coordinates and the boundary conditions imposed on the fields, appearing as the integration constants $c$'s and $d$'s (we refer to the appendix for all the relevant definitions).

First of all, notice that the solutions (\ref{Static-sol}) show singular behaviours both at $r=0$ and $r=1/H$.  In order to remove the singularity at the coordinates origin we will take $c_\infty = d_\infty = 0$ in the up coming discussion.

The general expressions for the energy momentum tensor, or rather the $tt$, $rr$, and $\theta\theta$ components were given in the appendix of~\cite{Mottola:2006ew}.  Notice that due to the spherical symmetry of the system the $\theta\theta$ and $\phi\phi$ components are the same.  Moreover, in de Sitter spacetime only the diagonal components of $T^A_{\mu\nu}$ survive, so the aforementioned three components is all we need.

As we want to keep the analysis as general as possible, we will simply expand the stress tensor in (inverse) powers of $(1 - H r)$, and study the coefficients of the singular parts.  Once these coefficients are found, we will demand that, in the most general way, they must not change in the course of a transition which may (should) change some of the set of coefficients $(b, b', c0, d0, c_1, d_1, c_2, d_2)$.  This requirement would then automatically imply that the divergencies (at the horizon) will automatically cancel, enabling us to look at the finite contributions.

Before embarking on this process, let us notice that, despite having removed the $r=0$ singularity from the auxiliary fields, in general there still may be terms with $1/r$ and $1/r^2$ singular behaviour in the stress tensor.  The simplest way to remove them is to take the coefficients $c_1$ and $c_2$ to be zero.  This will ensure that every component of (\ref{emtA}) is regular at the origin.

Hence, the next step is to expand the energy momentum tensor near the horizon.  In general there will be $1/(1 - H r)^2$, $1/(1 - H r)$ and $\ln(1 - H r)$ singular behaviours, with different coefficients for each of the three independent components.  However, it turns out that some of these coefficients are the same, or linearly dependent, and we can describe all of the divergencies in $T^A_{\mu\nu}$ by means of these three coefficients
\be\label{singcoeff}
\alpha &=& b' + 2 b' c2 + b' c_2^2 + 2 b d_2 + 2 b c_2 d_2 \, , \nonumber\\
\beta &=& 3 b' + 6 b' c_2 + b' c_2^2 + 6 b d_2 + 2 b c_2 d_2 \, , \\
\gamma &=& b' + 2 b' c_2 - b' c_2^2 + 2 b d_2 - 2 b c_2 d_2 \, . \nonumber
\ee

Thus, all the horizon divergencies of the stress tensor will disappear if these coefficients were simultaneously vanishing, which means that either
\be\label{nodiv1}
c_2 = 0 \qquad \textrm{and} \qquad d_2 = - b' / 2b \, ,
\ee
or, alternatively,
\be\label{nodiv2}
c_2 = -1 \qquad \textrm{and} \qquad d_2 = b' / 2b \, .
\ee
We note in passing that there is a typo in equation (4.28a) of~\cite{Mottola:2006ew}, as that equation should read, in their notation, $2b (d_H - p p') = b' (p^2 - 1)$. (This typo was confirmed through correspondence with Emil Mottola.)

Equations (\ref{nodiv1}) or (\ref{nodiv2}) would ensure a regular behaviour for the stress tensor everywhere at every time.  If one whishes to find less restrictive conditions, a different process for removing the divergencies must be specified.  Our proposal, as elucidated in the previous paragraphs, is to consider the energy momentum tensor components before and after a transition which acts on the numerical values of the parameters describing the auxiliary fields, as well as it changes the number of fields contributing to the anomaly.  If we attach a subscript $f$ for the final values taken by these coefficients, then the divergencies of $T^A_{\mu\nu}$ will cancel if
\be\label{nodivDyn}
\left\{
\begin{array}{l}
\alpha_f = \alpha \\
\beta_f = \beta \\
\gamma_f = \gamma
\end{array} \right.
\quad \Rar \quad
\left\{
\begin{array}{l}
c_{2f} = (b' - b'_f + 2 b' c_2 + 2 b d_2 - 2 b_f d_{2f}) / 2 b'_f \\
\\
d_{2f} = \pm \sqrt{(- b' + b'_f - 2 b' c_2 - 2 b d_2)^2 + 4 b'_f c_2 (- b' c_2 - 2 b d_2)} / 2 b_f
\end{array} \right. \, .
\ee

The conditions (\ref{nodivDyn}) guarantee that the horizon divergencies cancel out dynamically as a result of a quantum transition from one vacuum state to another, states which are described by the different boundary conditions assigned to the auxiliary fields.  Consequently, even though the stress tensor is still divergent when taken at a single time moment, the underlying idea is that what is observable is not the absolute value of the energy density (or pressure density), but only the difference between an initial and a final state.  As long as equation (\ref{nodivDyn}) is observed, there will be no divergences in the resulting (subtracted) $T^A_{\mu\nu}$. This process is in the spirit of standard QFT renormalization; we take our observable at a ``point of normalization" and subtract it from our observable at any other point.

Notice that this ``renormalisation prescription'' is not claimed to be the unique way to {\it define} the vacuum energy. Indeed, in a quantum field theory the ``point of normalisation" can be chosen quite arbitrary.  Physical results will depend on the specific input IR parameters at this ``point of normalisation". 
Our goal here is to demonstrate how our  proposal may in principle work 
 when  the cosmological constant   emerges from   IR   related physics  rather than from 
UV linked physics which is now a commonly accepted framework\cite{Copeland:2006wr}. 
 
We are now in the position to calculate the finite contribution to the energy and pressure density, obtained after the divergent parts have been removed by means of (\ref{nodivDyn}).  The result is ($\Delta = $ final state - initial state)
\be\label{deltaemtA}
\Delta {T^{A\, t}_{\:\ t}} = \Delta {T^{A\, r}_{\:\ r}} = \Delta {T^{A\, \theta}_{\:\ \theta}} = 6 H^4 \Delta b' \, ,
\ee
which does not depend on the coefficients $c_2$ and $d_2$, nor it does depend on $b$; the latter is easily understood since in de Sitter space $F=0$.  Slightly more surprising is the absence of the $c_2$ and $d_2$, but it is maybe ascribable to either the still high degree of symmetry that Static-dS possess, or to the simplifying assumptions made in deriving the solutions for $\varphi$ and $\phi$, or both, see next subsection for more comments on this.  

The result (\ref{deltaemtA}) is finite but numerically not comforting, as, although we were able to construct a well defined and working prescription for how to remove the divergences of the anomalous stress tensor, and we even obtained a finite remnant of this process, the numerical value of this finite part, if we aim at explaining the currently observed vacuum energy density, is way too small, as it is proportional to $H^4$ similar to (\ref{H^4}) rather than $\sim H$ we were hoping to obtain similar to (\ref{H}).

The reason that this energy density is so small may be that, despite the solutions for the auxiliary fields being kept quite general, requiring only that they (together with the stress tensor) do not diverge at the origin, in the final result (\ref{deltaemtA}) there is no sign of these fields. Therefore, no IR physics has penetrated into the final expression (\ref{deltaemtA}). As a result of this deficiency,  the contribution of the anomaly reduces once again to the local, second order in the curvature, geometrical terms, which are of order $H^4$, and therefore irrelevant at scales lower than the Planck one. We make few comments on this (negative) result in the next subsection.

\subsection{Perspectives. Future Developments. }\label{persp}
Below we consider the questions: first, why is our final expression (\ref{deltaemtA}) not sensitive to our input parameters ($c_2$ and $d_2$) which parametrise the IR physics in our computations? and second, what was the missing element in our present computations, and why we are not getting the desired scaling (\ref{H})?

\vspace{0.2cm}
\noindent
{\it  \underline{ More general solutions for $\varphi$ and $\psi$. }}

 The solutions we have provided for the auxiliary fields contain four integration constants each, but once one demands regularity of these solutions and of the stress tensor, two of these coefficients disappear.  Moreover, the constants $c_0$ and $d_0$ do not actually enter in any calculation, as the fields appear always differentiated.  This means that in fact, we are effectively working with only one parameter ($c_2$) for $\varphi$ and one ($d_2$) for $\psi$.  It is hence not surprising that, after imposing (\ref{nodivDyn}), there is no dependence on the boundary conditions in the final difference (\ref{deltaemtA}).  Consequently, it is necessary to generalise these solutions, and look for both ``time'' and ``space'' dependent auxiliary fields.
 
\vspace{0.2cm}
\noindent
{\it  \underline{Different metrics.}}

The metrics we have been working with so far, despite some of them not showing explicitely the full symmetries of de Sitter spacetime, still possess high degrees of symmetry.  Moreover, the assumption of idealised de Sitter space throughout our analysis is not justified by what we know about the history of the Universe.  Indeed, most of the physics we have described in section \ref{dealing} does not take place in a de Sitter background, but instead a radiation dominated FRW one.

It would be more realistic to include this ``transition'' of the metric as well, even though we may not expect any major deviation from the results obtained here since the curvature of the spacetime remains small. However, in this set up there is a real chance that the IR physics may penetrate into the final expression in contrast with our previous result (\ref{deltaemtA}) where the information about boundary conditions did not show up.  In addition, in this case it is not clear whether to expect the singularities to survive the singular coordinate transformation to the FRW set, as the latter possesses translation invariance which would preclude the appearance of a divergent stress tensor at the horizon.

\vspace{0.2cm}
\noindent
{\it  \underline{Coupling the auxiliary fields to SM particles.}}

 If we are to consider the auxiliary fields as actual physical degrees of freedom of the theory (notice that, despite their fourth order equations of motion, their spectra do not possess any negative norm states~\cite{Anderson:2002fk}), then we should allow for the possibility that these fields couple to other SM fields as well. In fact, the general arguments presented in~\cite{Schutzhold:2002pr} suggest that non-local (large distance)  interaction may play a crucial role in understanding the scaling (\ref{H}). The auxiliary fields  $\varphi$ and $\psi$ provide precisely this kind of non-locality if they are not integrated out. It is very likely that we are not finding the desired scaling (\ref{H}) just because the IR physics is not making it into the final expression  (\ref{deltaemtA}) as a result  of decoupling of auxiliary fields  $\varphi$ and $\psi$ from matter fields. In fact, a similar computations~\cite{Urban:2009wb} in a toy two dimensional model suggests that this kind of interactions could be a crucial element potentially leading to the desired scaling (\ref{H}). Hopefully, the lessons we learn from~\cite{Urban:2009wb} may teach us how to implement these ideas  in four dimensions \cite{Urban:2009vy}. 

\section{Conclusions}
Let us briefly summarise the main points of the analysis reported in this paper.

{\bf Gravity as a low energy EFT} -- We have adopted and reviewed the idea that Gravity is to be thought of as just a low energy EFT wherein the UV cutoff scale (be it the Planck mass or something else) should appear neither in the low energy effective Lagrangian, nor in the physical definitions of the low energy physical observables, such as the cosmological constant.  Such a theory must necessarily incorporate anomalies. In the Gravity case, the conformal anomaly can be described as an effective action thanks to two auxiliary scalar fields, whose dynamics can be shown to be relevant in certain specific setups (spacetimes with horizons).  In particular, the physical stress tensor in general blows up at such horizons, and a method to deal with these divergencies must be specified.

{\bf The cosmological constant problem as an IR effect} -- In this framework, it is obvious that the UV cutoff and its associated vacuum energy are doomed to vanish once a definition of renormalisation is provided, possibly leaving behind a residual energy density whose scale will be given by the IR scale described by the auxiliary fields.  We have defined the physical vacuum energy as a dynamical result of a cancellation that relates two sides of a ``phase transition'' in which the fields participating to the anomaly change their properties and therefore their contribution to the anomalous stress tensor.

{\bf The scale of the vacuum energy} -- Following the general path just proposed, we have employed a specific, simplified, solution for the auxiliary fields in static de Sitter spacetime, and we have shown how it is possible to cancel the divergencies by imposing certain conditions on the way the BC's for these fields change, and we have also found that there indeed is a finite, non-vanishing, relic vacuum energy.  Unfortunately, this vacuum energy is proportional to $H^4$ which makes it unviable for cosmological applications, and we have shown why, in this oversimplified setup, the result could not possibily be different (in short, the QFT IR scale does not appear in the final result, leaving only $H$ as the physical scale).  However, this is not the only possibility, as we expect that, once more general solutions for the auxiliary fields are found, a mixing between curvature and the IR scale of the effective QFT will take place.

\acknowledgments
ARZ would like to thank Grisha Volovik for correspondence, J.~Bjorken for sharing his ideas on vacuum energy, and Emil Mottola and   Alexei Starobinsky for discussions. FU would like to thank R.~Woodard and P.~Moyassari for helpful conversations, and the Kavli Institute for Theoretical Physics China in Beijing for kind hospitality and support while this work was being finalised. This research was supported in part by the Natural Sciences and Engineering Research Council of Canada.

\section*{APPENDIX}
In this appendix we list the coordinate systems, with relative coordinate transformations, and the resulting equations of motion, with some solutions for the auxiliary fields, including those referred to in the text.

The FRW-dS system is given in terms of the coordinates $(\tau,\rho,\theta,\phi)$ as
\be\label{FRW-dS}
ds^2 = d\tau^2 - e^{2 H \tau} \left( d\rho^2 + \rho^2 d\Omega^2 \right) \, ,
\ee
with $d\Omega$ meaning $d\theta^2 + \sin^2\theta d\phi^2$.  This can be readily transformed into a conformally flat system by means of $\tau = - \ln(- H \eta) / H$, which maps the FRW-dS coordinates into the CF-dS set $(\eta,\rho,\theta,\phi)$.  The resulting line element is
\be\label{CF-dS}
ds^2 = \frac{1}{H^2 \eta^2} \left[ d\eta^2 - d\rho^2 - \rho^2 d\Omega^2 \right] \, .
\ee
These two systems are almost equivalent, as no mixing occurs between different coordinates.  This will be true of the next three sets as well, the first of which (Tortoise-dS) is obtained by performing the following consecutive transformations
\be\label{mixtransf}
\left\{
\begin{array}{l}
\rho = \omega / H \sinh \chi \\
\eta = \omega / H \cosh \chi
\end{array} \right.
\qquad \textrm{and} \qquad
\left\{
\begin{array}{l}
\omega = e^{- H t} \\
\chi = H r^*
\end{array} \right. \, ,
\ee
that returns the set $(t,r^*,\theta,\phi)$ and a line element of the form
\be\label{Tortoise-dS}
ds^2 = \sech^2(H r^*) \left[ dt^2 - {dr^*}^2 - \sinh^2(H r^*) / H^2 d\Omega^2 \right] \, .
\ee
The proper Static-dS system $(t,r,\theta,\phi)$ immediately follows from
\be\label{tortoise}
r^* = \frac{1}{2 H} \ln \frac{1 + H r}{1 - H r} \, ,
\ee
with a corresponding familiar line element
\be\label{Static-dSapp}
ds^2 = (1 - H^2 r^2) dt^2 - \frac{1}{1 - H^2 r^2} dr^2 - r^2 d\Omega^2 \, .
\ee
The last transformation is $t = T / H$ together with $r = R / H / \sqrt{1 + R^2}$, taking the Static-dS set into $(T,R,\theta,\phi)$ and transforming the line element to the ``Rindler-dS'' form
\be\label{Rindler-dS}
ds^2 = \frac{1}{(1 + R^2)H^2} \left[ dT^2 - \frac{1}{1 + R^2} dR^2 - R^2 d\Omega^2 \right] \, .
\ee

The equation of motion for the auxiliary fields were given in (\ref{auxphi}, \ref{auxpsi}).  The fourth order differential operator $\Delta_4$ in our five different coordinate systems becomes
\be
\Delta_4 = \left\{
\begin{array}{ll}
\partial^4_\tau + 6 H \partial^3_\tau + 11 H^2 \partial^2_\tau + 6 H^3 \partial_\tau + e^{- 4 H \tau} \partial^4_\rho - 2 e^{- 2 H \tau} \partial^2_\rho \left( \partial^2_\tau + H \partial_\tau \right) & \textrm{(FRW-dS)} \label{diff-FRW-dS} \\
&\\
&\\
\eta^4 H^4 \left[ \partial^4_\eta -2 \partial^2_\eta \partial^2_\rho + \partial^4_\rho \right] & \textrm{(CF-dS)} \label{diff-CF-dS} \\
&\\
&\\
\cosh^4(H r^*) \left[ \partial^4_t - 4 H^2 \partial^2_t + \partial^4_{r^*} + 4 H^2 \partial^2_{r^*} - 2 \partial^2_t \partial^2_{r^*} + 4 H \coth(H r^*) \left( \partial^3_{r^*} - \partial_{r^*} \partial^2_t \right) \right] & \textrm{(Tortoise-dS)} \label{diff-Tortoise-dS} \\
&\\
& \qquad\qquad\qquad\quad , \\
\partial^4_t - 2 \partial^2_t \partial^2_r - \frac{4}{r} \partial^2_t \partial_r + (1 - H^2 r^2)^2 \partial^4_r + & \\
& \textrm{(Static-dS)} \label{diff-Static-dS} \\
+ \frac{4}{r} (1 - H^2 r^2) (1 - 3 H^2 r^2) \partial^3_r - 4 H^2 (7 - 9 H^2 r^2) \partial^2_r - \frac{8}{r} H^2 (1 - 3 H^2 r^2) \partial_r & \\
&\\
&\\
H^4 (1 + R^2)^2 \left[ \partial^4_T - 4 \partial^2_T - 2 (1 + R^2) \partial^2_T \partial^2_R - \frac{2}{R} (2 + 3 R^2) \partial^2_T \partial_R + \right. & \\
& \textrm{(Rindler-dS)} \label{diff-Rindler-dS} \\
\left. + (1 + R^2)^2 \partial^4_R + \frac{2}{R} (2 + 7 R^2 + 5 R^4) \partial^3_R + (20 + 23 R^2) \partial^2_R + \frac{1}{R} (4 + 9 R^2) \partial_R \right] &
\end{array} \right.
\ee
where we have restricted the coordinate dependence of both fields to the ``radial'' and ``time'' coordinates.

These equations are fairly complicated, and so far we were able to obtain a general solution only for the conformally flat spacetime can be found (but in this case, see~\cite{Mottola:2006ew}, all the dependence of the anomalous stress tensor on the auxiliary fields drops out).  However, it is not impossible to solve these fourth order differential equations if we drop, for instance, the ``time'' dependence from the fields $\varphi$ and $\psi$.  Focussing on the last three spacetime coordinatisations, where the presence of the horizon is explicit, we can simplify the Euler-Lagrange equations for the auxiliary fields as follows
\be
\cosh^4(H r^*) [ \partial^4_{r^*} + 4 H \coth(H r^*) \partial^3_{r^*} + 4 H^2 \partial^2_{r^*} &\!\!\! ] \left\{
\begin{array}{c}
\varphi \\
\psi
\end{array}
\right\} &\! = \left\{ \!
\begin{array}{c}
12 H^4 \\
0
\end{array} \!
\right\} \qquad \textrm{(Tortoise-dS)} \, , \label{Tortoise-eom} \\
\nonumber\\
\, [ \, (1 - H^2 r^2)^2 \partial^4_r + \frac{4}{r} (1 - H^2 r^2) (1 - 3 H^2 r^2) \partial^3_r && \nonumber\\
- 4 H^2 (7 - 9 H^2 r^2) \partial^2_r - \frac{8}{r} H^2 (1 - 3 H^2 r^2) \partial_r &\!\!\! ] \left\{
\begin{array}{c}
\varphi \\
\psi
\end{array}
\right\} &\! = \left\{ \!
\begin{array}{c}
12 H^4 \\
0
\end{array} \!
\right\} \qquad \textrm{(Static-dS)} \, , \label{Static-eom} \\
\nonumber\\
H^4 (1 + R^2)^2 \, [ \, (1 + R^2)^2 \partial^4_R + \frac{2}{R} (2 + 7 R^2 + 5 R^4) \partial^3_R && \nonumber\\
+ (20 + 23 R^2) \partial^2_R + \frac{1}{R} (4 + 9 R^2) \partial_R &\!\!\! ] \left\{
\begin{array}{c}
\varphi \\
\psi
\end{array}
\right\} &\! = \left\{ \!
\begin{array}{c}
12 H^4 \\
0
\end{array} \!
\right\} \qquad \textrm{(Rindler-dS)} \, . \label{Rindler-eom}
\ee

The simplest of these equations are of course the ones in the Toirtoise-dS system, equation (\ref{Tortoise-eom}).  Once a solution for these equations have been found, it is immediate to transform them into the Static-dS and Rindler-dS sets.  The general solution to equations (\ref{Tortoise-eom}) can be written as
\be
&\left\{
\begin{array}{l}
\varphi(r^*) = c_\infty \ctgh(H r^*) + c_0 - 2 c_1 H r^* - 2 c_2 \ctgh(H r^*) H r^* + 2 \ln\left[ \sech(H r^*) \right] \\
\\
\psi(r^*) = d_\infty \ctgh(H r^*) + d_0 - 2 d_1 H r^* - 2 d_2 \ctgh(H r^*) H r^*
\end{array} \right. & \textrm{(Tortoise-dS)} \label{Tortoise-sol} \, , \\
&\nonumber\\
&\nonumber\\
&\left\{
\begin{array}{l}
\varphi(r) = \frac{c_\infty}{H r} + c_0 + c_1 \ln\frac{1 - H r}{1 + H r} + \frac{c_2}{H r} \ln\frac{1 - H r}{1 + H r} + \ln(1 - H^2 r^2) \\
\\
\psi(r) = \frac{d_\infty}{H r} + d_0 + d_1 \ln\frac{1 - H r}{1 + H r} + \frac{d_2}{H r} \ln\frac{1 - H r}{1 + H r}
\end{array} \right. & \textrm{(Static-dS)} \label{Static-sol} \, , \\
&\nonumber\\
&\nonumber\\
&\left\{
\begin{array}{l}
\varphi(R) = \frac{c_\infty}{R} \sqrt{1 + R^2} + c_0 + c_1 \ln\frac{\sqrt{1 + R^2} - R}{\sqrt{1 + R^2} + R} + \frac{c_2}{R} \ln\frac{\sqrt{1 + R^2} - R}{\sqrt{1 + R^2} + R} - \ln(1 + R^2) \\
\\
\psi(R) = \frac{d_\infty}{R} \sqrt{1 + R^2} + d_0 + d_1 \ln\frac{\sqrt{1 + R^2} - R}{\sqrt{1 + R^2} + R} + \frac{d_2}{R} \ln\frac{\sqrt{1 + R^2} - R}{\sqrt{1 + R^2} + R}
\end{array} \right. & \textrm{(Rindler-dS)} \label{Rindler-sol} \, ,
\ee
where the integration constants have been chosen in such a way that the coefficients are all 1 in the Static-dS system.  The solutions for the Static-dS spacetime parametrisation were alrady given in~\cite{Mottola:2006ew}, and coincide with those given here.  Notice that there needs not to be perfect symmetry between the integration constants $c$'s and $d$'s, as the inhomogeneous solution will in general break this parallelism.

It is fairly straightforward to write down the solutions for the FRW-dS and CF-dS spaces, which we do for completeness.  Again, assuming only ``radial'' dependence we obtain the following equations
\be
\eta^4 H^4 \partial^4_\rho &\!\!\! \left\{
\begin{array}{c}
\varphi \\
\psi
\end{array}
\right\} &\! = \left\{ \!
\begin{array}{c}
12 H^4 \\
0
\end{array} \!
\right\} \qquad\qquad \textrm{(FRW-dS)} \, , \label{FRW-eom} \\
\nonumber\\
e^{- 4 H \tau} \partial^4_\rho &\!\!\! \left\{
\begin{array}{c}
\varphi \\
\psi
\end{array}
\right\} &\! = \left\{ \!
\begin{array}{c}
12 H^4 \\
0
\end{array} \!
\right\} \qquad\qquad \textrm{(CF-dS)} \, , \label{CF-eom}
\ee

Time dependence can be easily introduced in the CF-dS system, but as already noticed, in this (highly symmetric) spacetime, all the dependence of the anomalous stress tensor from the auxiliary fields drops out.  We write down the general solutions to equations (\ref{FRW-eom}) and (\ref{CF-eom}), differing only for the inhomogeneous term, for reference.
\be
&\left\{
\begin{array}{l}
\varphi(\rho) = k_0 + k_1 \rho + k_2 \rho^2 + k_3 \rho^3 + \frac{\rho^4}{2 \eta^4} \\
\\
\psi(\rho) = k'_0 + k'_1 \rho + k'_2 \rho^2 + k'_3 \rho^3
\end{array} \right. & \qquad \textrm{(FRW-dS)} \label{FRW-sol} \, , \\
&\nonumber\\
&\left\{
\begin{array}{l}
\varphi(\rho) = k_0 + k_1 \rho + k_2 \rho^2 + k_3 \rho^3 + \frac{H^4 \rho^4}{2} e^{- 4 H \tau} \\
\\
\psi(\rho) = k'_0 + k'_1 \rho + k'_2 \rho^2 + k'_3 \rho^3
\end{array} \right. & \qquad \textrm{(CF-dS)} \label{CF-sol} \, .
\ee

Had we allowed for ``time'' dependence instead, we would have obtained solutions that, especially for the CF-dS system, look very similar to those shown above, with the r\^ole of ``radius'' and ``time'' interchanged.

One last comment before proceeding further to the computation of the energy momentum tensor.  The inhomogeneous solution for $\varphi$ always corresponds to the so called Bunch-Davies vacuum, which is the solution which brings every coordinate system back to the conformally flat case, see~\cite{Mottola:2006ew}.  As long as this solution is employed (that is, with all the other constants taken to be zero), by definition all the dependence on the auxiliary fields disappear, and there is no reason to expect this new description of the anomalous action to have any effect at all that is not entirely describable in the standard formalism.  This is exactly what happens in the analysis of~\cite{Koksma:2008jn} (see also~\cite{Dappiaggi:2008mm}), where their results apply only to the simplest case where the Bunch-Davies vacuum is realised, and no effects of the boundary conditions can show up in the resulting vacuum energy.


\end{document}